\documentclass[preprint]{revtex4}
\usepackage{graphicx,bm}
\usepackage{amssymb,color,amsmath}
\def\II{\mathcal{I}}

\def\tr{\textrm{tr }}

\def\p{{\bm{p}}}
\def\q{{\bm{q}}}
\newcommand{\be}{\begin{equation}}
\newcommand{\ee}{\end{equation}}

\newcommand{\x}{{\mathbf x}}

\newcommand{\vct}[1]{\ensuremath\mbox{\boldmath$ #1 $}}
\newcommand{\Vxi}{\vct \xi}
\newcommand{\Veta}{\vct \eta}

\def\ro{\hat{\rho}}
\begin{document}
\title{Semiclassical theory for small displacements}
\author {Eduardo Zambrano\footnote{zambrano@cbpf.br} and Alfredo M. Ozorio de Almeida\footnote{ozorio@cbpf.br}}
\address{Centro Brasileiro de Pesquisas Fisicas,
Rua Xavier Sigaud 150, 22290-180, Rio de Janeiro, R.J., Brazil}
\date{\today}
\date{December 9, 2008}
\begin{abstract}
Characteristic functions contain complete information about all the moments of a classical distribution
and the same holds for the Fourier transform of the Wigner function: a {\it quantum characteristic function},
or the chord function. However, knowledge of a finite number of moments does not allow for accurate
determination of the chord function. For pure states this provides the overlap of the state with
all its possible rigid translations (or displacements). We here present a semiclassical approximation
of the chord function for large Bohr-quantized states, which is accurate right up to a caustic,
beyond which the chord function becomes evanescent. It is verified to pick out blind spots,
which are displacements for zero overlaps. These occur even for translations within a Planck area
of the origin. We derive a simple approximation for the closest blind spots, depending on the 
Schr\"odinger covariance matrix, which is verified for Bohr-quantized states.
\end{abstract}

\maketitle
\section{Introduction}

Experiments in quantum optics (see e.g. \cite{Leonhardt}),
atom traps, or other quickly developing technologies are rapidly realizing 
the promiss of a manipulative quantum mechanics.
The object of developing the field of quantum information has led to a refinement of techniques, 
such that the interference between the wave functions of single atoms, 
or single modes of an optical cavity, can be measured.
So far, experiments have mainly been realized for very simple states,
but theory can anticipate a future where more delicate states, such as the
superposition of many coherent states or the excited state of an anharmonic oscillator
may be made to interfere. Experimental work with Rydberg atoms already point in this direction \cite{Haroche}. 

A typical interference experiment superposes two modified copies
of the same initial state. 
For instance, in quantum optics, it is easy to achieve
the unitary transformation that corresponds to a uniform
phase space translation (or displacement) of the phase space variables $\x=(p,q)$. 
This translated state can then interfere with the original state.
In general, the unitary {\it translation operator}
\begin{equation}
\hat{T}_{\Vxi} = \exp{\left[\frac{i}{\hbar}(\Vxi\wedge \hat{\x})\right]}=
\exp{\left[\frac{i}{\hbar}(\Vxi_p\cdot \hat{q}-\Vxi_q\cdot \hat{p})\right]} \ ,
\end{equation}
acts on the state $|\psi\rangle$ to produce the new state
$|\psi_{\Vxi}\rangle=\hat{T}_{\Vxi}|\psi\rangle$ in strict
correspondence to the classical translation, $\x \mapsto \x + \Vxi$,
by the {\it chord} $\Vxi$.
\footnote{ In the optical context $\hat{T}_{\Vxi}$ is
usually referred to as the {\it displacement operator} and is
expressed in terms of creation and annihilation operators for the
harmonic oscillator. This is inconvenient for semiclassical
analysis.}
Thus, given an arbitrary superposition of a state and its translation,
$a|\psi\rangle+b|\psi_{\Vxi}\rangle$, with $|a|^2+|b|^2=1$,
the probability that this is measured to be in the untranslated state
is $|a+ b\langle\psi|\psi_{\Vxi}\rangle|^2$.

Evidently, measurements of such probabilities (through
repeated preparation) supply detailed quantum information
concerning these initial states.
Better still, the full set of possible overlaps defines the
complete phase space representation,
\be
\chi (\Vxi)= \langle\psi|\hat{T}_{-\Vxi}|\psi\rangle.
\label{chitrans}
\ee
This is known as the {\it chord function} \cite{OzReport}, 
as one of the {\it quantum characteristic functions} of quantum optics \cite{Leonhardt}
(or the {\it Weyl function} as in \cite{Chountasis}),
which is the Fourier transform of the Wigner function \cite{Wigner}:
\begin{equation}
 \chi(\Vxi) = \int d\x \;W(\x)
\exp{\left\{\frac{i}{\hbar}(\Vxi\wedge \x)\right\}} \ .
\label{chift}
\end{equation}
The latter can be redefined, following Royer \cite{OzReport,Royer}, as
\be
W(\x)=\frac{1}{(\pi\hbar)}\langle\psi|\hat{R}_{\x}|\psi\rangle,
\label{Wrefl}
\ee
where $\hat{R}_{\x}$, the Fourier transform of the translation operators,
corresponds classically to the phase space reflection through the {\it reflection centre} $\x$,
i.e. $\x_0 \mapsto 2\x - \x_0$.
 
\begin{figure}[htb!]
\centering
\includegraphics[width=12cm]{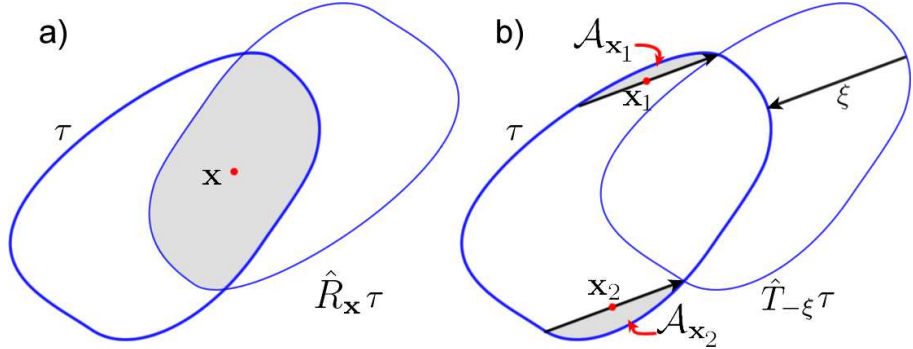}
\caption{\label{pe}
Classical reflection a) and translation b) of a Bohr-quantized curve. Semiclassically, the intersection between the original and the transformed curve determines the stationary points and the semiclassical phase is given by the area bounded by them (shadow area).}
\end{figure}

Thus, both the Wigner function and the chord function provide complete information
about states, by telling us how they respond to certain continuous sets of quantum manipulations.
These translation and reflection operators act on Hilbert space in close correspondence
to classical phase space translations and reflections. 
Hence, their action on excited states of anharmonic oscillators 
should correspond closely to the translations and reflections
of the Bohr-quantized curves on which they are classically supported.
This is well verified in the context of the semiclassical theories for the Wigner function \cite{Berry77}
and the chord function \cite{OVS, ZO}. The Wigner function, $W(\x)$, oscillates with a non-negligible amplitude
for all reflection centres, $\x$, such that the quantized curve and its reflection intersect transversely, 
as shown in Fig. \ref{pe}$a$.
Likewise, all translation chords, $\Vxi$, for which the translated curve intersects the quantized curve 
transversely, are in the region where the chord function, $\chi(\Vxi)$, displays sizeable oscillations.
Since the quantized curve is closed, it intersects its reflection around $\x$, or its translation by $\Vxi$,
at an even number of points. In both cases, it is the area between the pair of curves joining two intersections
that determines the phase of the oscillations of the Wigner and the chord function respectively, the shadow areas in Fig. \ref{pe}.   

However, it turns out that in both cases the classical region associated to the quantized curve
lies on caustics where a simple semiclassical theory breaks down. 
In the case of the Wigner function, this caustic is the locus of reflection centres in the neighbourhood 
of the quantized curve itself, that is, as the centre, $\x$, approaches the curve, 
pairs of intersections with the reflected curve coalesce. 
This caustic is generic and was already dealt with in Berry's original treatment \cite{Berry77}.
In contrast, the classical region in the space of translation chords is the neighbourhood of the origin,
whatever the shape of the quantized curve. This is highly nongeneric, because all points on the quantized
curve {\it intersect} the translated curve in the limit as $\Vxi \rightarrow 0$, i.e. both curves coincide. 
Another (nonclassical) caustic, associated to the longest chords for which the translated curve still touches
the Bohr-quantized curve, was already treated semiclassically in \cite{ZO} (along with the corresponding
Wigner caustic). Thus, the present theory, joining the short chord region to the oscillatory region
completes the general picture for the chord function of a Bohr-quantized state with a single degree of freedom. 

Notwithstanding the difficulty of including the neighbourhood of the origin of chords in a semiclassical theory,
this region encodes a rich store of information concerning the quantum state. On the one hand, the derivatives
of the chord function, evaluated at the origin, specify all the moments of position and momentum and their products.
For a Bohr-quantized state, the moments can be identified with classical averages over the corresponding quantized curve,
which is one justification for considering a Planck area surrounding the chord origin to be a classical region.
In contrast, one finds points of complete orthogonality between the state and its translation
within this same neighbourhood, which shows that classical correspondence cannot be pushed too far. 
In the general case where the state has no reflection symmetry,
orthogonality occurs for isolated points. A theoretical treatment for the pattern of these special chords, 
named {\it blind spots}, was presented for arbitrary superpositions of coherent, or squeezed states in \cite{Blind}.
One of the objectives here is to extend this analysis to Bohr-quantized states.

Our starting point is a simple integral formula for the chord function that is only valid for small chords.
This was presented in \cite{OVS}, but we here provide a fresh rederivation in section 2 and then go on to
show that it leads to classical expressions for the moments. Of course, the knowledge of all the moments
provides a Taylor series for the chord function, but its finite polynomial approximations 
cannot be joined smoothly to the oscillatory region, which is well described by the standard semiclassical
theory in \cite{OVS, ZO}. In section 3, we establish the presence of blind spots in a neighborhood of the origin for any extended state. We present, in section 4, an interpolation that bridges the two regimes of the chord function (small and long chords).
The full semiclassical theory is compared numerically with the exact result in section 5, 
for an example of a Fock state that is subjected to an unitary transformation which breaks its reflection symmetry. Finally, we discuss our results in section 6.

\section{Small chords and moments of position and momentum}

Consider a general semiclassical WKB state associated to the one-dimensional classical manifold defined by $S(q,I=\II)$ \cite{Berry77,VanVleck,Maslov}, 
\be
\langle q|\psi_\mathcal{I}\rangle=
N\sum_j\left|\frac{\partial^2S_j(p,q)}{\partial q\partial I}\right|^{-\frac12}
e^{\frac{i}\hbar S_j(q,I)+i\beta_j},
\ee
where $S$ is the generating function of the canonical transformation between the cartesian variables and the \emph{action-angle} variables, namely $\x=(p,q)\mapsto(I,\theta)$, the index $j$ enumerates the branches of $S$ 
and $\beta$ is the Maslov correction. The chord function for a WKB state, 
obtained by translating this state and taking its overlap according to (\ref{chitrans}),
is given by the superposition 
\begin{equation}
\chi_{w}(\Vxi)=\sum_{jk} \chi_{jk}(\Vxi),
\label{super}
\end{equation}
where the terms $\chi_{jk}$ are given by
\be
\chi_{jk}(\Vxi)=N^2
\int dQ	
\left|\frac{\partial^2S_j}{\partial q\partial I}\left(Q_+\right)
      \frac{\partial^2S_k}{\partial q\partial I}\left(Q_-\right)\right|^{-\frac12}
e^{\frac{i}{\hbar}
    \left[S_j(Q_+)-S_k(Q_-)-\xi_p Q\right]
    +(\beta_j-\beta_k)},
\label{X-WKB}
\ee
with $Q_\pm\equiv Q\pm\frac{\xi_q}{2}$. Thus, the oscillatory semiclassical form for each branch of 
the chord function results from the stationary phase evaluation of this integral, 
to be discussed in section 3, but we are here concerned with the limit where the translation
is so small that the phase between stationary points does not rise above Planck's constant.

Due to the symplectic invariance of the chord function \cite{OzReport}, 
we may choose $\Vxi$ to be parallel to the vertical axis, without loss of generality. 
In this case, $Q_+ = Q_- = Q$, so that the phase difference between the top and bottom branches
of the curve is just the curve area as a function of $Q$, which is large and not stationary.
Thus, the neglect of these terms leaves only the `diagonal' terms in \eqref{super}, which are given by
\be
\sum_j\chi_{jj}(\Vxi)=N^2
\sum_j\int dQ\left|\frac{\partial\theta_j}{\partial q}\right|e^{\frac{i}\hbar\xi_p Q}
=N^2\int_0^{2\pi}\,d\theta\, e^{-i\xi_p Q(\theta)/\hbar}.
\label{diagonalterms}
\ee
Again, making use of symplectic invariance, the right-hand expression can be identified with
the semiclassical approximation of the chord function for short chords, introduced in \cite{OVS}:
\be
\chi(\Vxi)\simeq\int_0^{2\pi}\,\frac{d\theta}{2\pi}e^{i\x(\theta)\wedge\Vxi/\hbar}
=\int_0^{2\pi}\,\frac{d\theta}{2\pi}e^{i[p(\theta)\xi_q-q(\theta)\xi_p]/\hbar}.
\label{chi-OVS}
\ee
This approximation assumes that the classical curve is specified by \emph{action-angle} variables, 
that is, $I(p,q)=\II$ and, conversely, $\x(\theta)\equiv(p(\theta),q(\theta))$. 

The formula \eqref{chi-OVS} holds for any choice of the direction of the small chord $\Vxi$ \cite{OVS}.
It describes the purely classical features of the state, in as much as it is the exact Fourier transform 
of the `classical approximation' of the Wigner function, $W(\x) =  \delta[I(p,q)=\II]/2\pi$, 
proposed by Berry \cite{Berry77}. However, it is more precise to consider this form of the Wigner function
as a rash extrapolation of the correct form of the small chord approximation to arbitrarily large chords.
Indeed, it will be here shown that the small chord version encodes quantum orthogonalities, as well as classical moments. 

The definition of the chord function \eqref{chitrans} allows us to calculate the statistical moments of $\hat\p$ and $\hat\q$ in the form of derivatives of the chord function, i.e. explicitly
\be
\langle\hat\p^n\rangle=\tr\,\hat\p^n\ro=(-i\hbar)^{n}\left.\frac{\partial^n\chi}{\partial\xi_q^n}\right|_{\Vxi=0}
\quad\quad\textrm{ and }\quad\quad
\langle\hat\q^n\rangle=\tr\,\hat\q^n\ro=(i\hbar)^{n}\left.\frac{\partial^n\chi}{\partial\xi_p^n}\right|_{\Vxi=0}.
\label{deri-moment}
\ee
Conversely, if we know all the moments, then we know the chord function, 
because the expansion in a Taylor series of the chord function is
\be
\chi(\Vxi)=
\sum_{n=0}^{\infty}\frac1{n!}\sum_{k=0}^{n}\frac{(-1)^{k}}{(i\hbar)^n}\left(\hspace{-.2cm}\begin{array}{c}n\\n-k\end{array}\hspace{-.2cm}\right)
\left\langle\mathcal{M}\left(\hat\q^{n-k}\hat\p^{k}\right)\right\rangle\hspace{.1cm}\xi_q^k\xi_p^{n-k},
\label{X-exp-momentdis}
\ee
where,
\be
\mathcal{M}\left(\hat\q^{n}\hat\p^{k}\right)=
\frac{1}{n+k}\sum_{P_{nk}}\hat\q^{n}\hat\p^{k}
\label{permutations}
\ee
and $P_{nk}$ are all possible permutations of products of $q^n$ and $p^k$. 
The equation \eqref{permutations} corresponds to the symmetrization of the product $\q^n\p^k$, 
being the important feature of the Weyl symbols, which guarantees the symplectic invariance of the chord function.
\par

According to \eqref{deri-moment} and \eqref{chi-OVS}, for a given parameterization, the moments are obtained by evaluating integrals of the form
\be
\langle \q^n\rangle\sim
\int_0^{2\pi}\frac{d\theta}{2\pi}\,[\q(\theta)]^n
\textrm{\,\, and }\,\,
\langle \p^n\rangle\sim
\int_0^{2\pi}\frac{d\theta}{2\pi}\,[\p(\theta)]^n.
\label{moment-classic}
\ee
These formulas correspond to the classical expected values of powers of position and momentum. 
They can also be obtained directly from the classical approximation of the Wigner function,
even though this not so satisfactory in other respects. 

\section{Nodal lines and blind spots}

The chord function \eqref{chift} is the Fourier transform of the real Wigner function and is in general complex.
Indeed, the fact that it represents a Hermitian operator only implies the constraint, 
$\chi(-\Vxi) = \chi(\Vxi)^*$, where the asterix denotes complex conjugation.
On the other hand, the cosine and the sine transforms, $c(\Vxi)$ and $s(\Vxi)$ of the Wigner function are real, 
i.e. the real and imaginary part of $\chi(\Vxi)$, respectively. In line with the definition \eqref{chitrans},
we may construct the hermitian operators 
\begin{equation}
\hat c_{\Vxi} = \frac{\hat{T}_{\Vxi} + \hat{T}_{-\Vxi}}2=\cos\left(\frac{\xi\wedge\x}\hbar\right)
\quad\textrm{ and }\quad
\hat s_{\Vxi} = \frac{\hat{T}_{\Vxi} - \hat{T}_{-\Vxi}}{2i}\sin\left(\frac{\xi\wedge\x}\hbar\right),
\end{equation} so that 
$c(\Vxi) = \langle\psi|\hat c_{\Vxi}|\psi\rangle$  and
$s(\Vxi) = \langle\psi|\hat s_{\Vxi}|\psi\rangle$.
\footnote{ An alternative interpretation is to consider $c(\Vxi)$ and $s(\Vxi)$ to be the chord functions for appropriately symmetrized states \cite{Blind}.}. So these are generalizations of the potentials $\cos(k q)$ and $\sin (k q)$ of cold atoms illuminated by standing waves from lasers \cite{Harper,Aubry}, where $k$ is the wave vector of the laser and $q$ the position coordinate. 
In the case when the state $|\psi\rangle$ has a centre of symmetry
(i.e. there exists a centre, $\x$, such that $\hat R_{\x}|\psi\rangle = \pm |\psi\rangle$)
then $s(\Vxi)=0$, so that $\chi(\Vxi)=c(\Vxi)$ \cite{OVS, Blind}. 

In the general case where there is no reflection symmetry, an intersection of a nodal line
of $c(\Vxi)$ with a nodal line of $s(\Vxi)$ defines a {\it blind spot} \cite{Blind}, at which the translated state, $|\psi_{\Vxi}\rangle$, becomes orthogonal to $|\psi\rangle$ (some examples are illustrated in Figs. \ref{fock5} and \ref{fig-chi}).
Because $\chi(0)= 1$, the origin lies on a nodal line of $s(\Vxi)$.
In a neighborhood of the origin, the chord function may be approximated by
\begin{eqnarray}
\chi(\Vxi)&=&\langle\hat T_{-\Vxi}\rangle
\simeq
\left\langle1-\frac{i}{\hbar}\Vxi\wedge\hat \x-\frac{1}{\hbar^2}(\Vxi\wedge\hat \x)^2+\cdots\right\rangle
\\&=&
1-\frac{i}\hbar\Vxi\wedge\langle\hat \x\rangle-\frac{1}{\hbar^2}\langle(\Vxi\wedge\hat \x)^2\rangle+\cdots
\label{aproxto2}
\end{eqnarray}
Thus, the nodal line of $s(\Vxi)$ crossing the origin is locally parallel to the direction of $\langle\hat\x\rangle$.

On the other hand, because the origin is a local maximum of the chord function, 
the nodal lines of $c(\Vxi)$ for small chords avoid the origin. 
It follows from \eqref{aproxto2} that the  closest nodal line surrounding the origin is given approximately by
\begin{equation}
\langle(\Vxi\wedge\hat \x)^2\rangle = \langle\hat q^2\rangle\xi_p^2+\langle \hat p^2\rangle\xi_q^2 -
2\langle \hat q\hat p+\hat p \hat q\rangle\xi_q\xi_p= \Vxi \mathbf K \Vxi = \hbar^2,
\label{ellipse}
\end{equation} 
if we neglect higher order terms. This positive quadratic form is defined in terms of the Schr\"odinger covariance matrix 
\cite{Schr}, $\mathbf K$, which establishes the extent of the state in phase space, 
that is, $\det \mathbf K \geq (2\pi\hbar)^2$ is just the symplectically invariant version of the uncertainty principle.
The nodal line of $c(\Vxi)$ is thus approximated by the ellipse \eqref{ellipse} and the closest blind spot 
lies near the tip of the diameter parallel to $\langle\hat\x\rangle$. 
It is important to note that the present estimate for the pair of {\it closest blind spots}
depends only on the first and second order moments.
In the case of a quantized curve treated here, it will verified that the qualitative features of the nodal lines are explained by this simple approximation. However, the nodal lines of $c(\Vxi)$ may show marked influence of the higher order moments, in the case of a superposition of coherent states \cite{Blind}. 

The highly excited Bohr-quantized states appropriate for semiclassical treatment 
have a covariance matrix that is well described by the classical averages discussed in the previous section,
such that $\det \mathbf K \gg (2\pi\hbar)^2$. 
Thus, the ellipse \eqref{ellipse} lies in the deep interior of a neighbourhood of the origin 
with $\sqrt{\hbar}$ linear dimensions.
This is in line with the discussion in \cite{Blind}: Notwithstanding the delicate quantum nature
of blind spots, they can be found in the `classical' neighbourhood of the origin and they
are precisely determined by classical features. This apparent paradox is resolved by
the reciprocal relation between large and small scales of pure states in phase space,
that follows from the universal invariance of the intensity of the chord function for pure states
with respect to Fourier transformation \cite{Chountasis, OVS}: 
\begin{equation}
 |\chi(\Vxi)|^2 = \frac{1}{(2\pi\hbar)}\int d\Veta \; |\chi(\Veta)|^2
\exp{\left\{\frac{i}{\hbar}(\Vxi\wedge \Veta)\right\}} \ .
\label{invariance}
\end{equation}
So far, we have only estimated the closest blind spots to the origin. It is hopeless to pursue the Taylor
expansion \eqref{aproxto2} any further to find further orthogonalities. On the other hand, 
the real and imaginary parts of the short chord approximation \eqref{chi-OVS}
have many nodal lines in the region where it holds, in the case of a big Bohr-quantized state.
In the following section, an interpolation formula is presented that allows for a uniform description,
which is also valid in the outer oscillatory region, where the chord function is evaluated by stationary phase.

\section{Joining the long and short chord regimes}

In section 2 we rederived the `classical' approximation \eqref{chi-OVS},
which holds in the neighborhood of the origin, 
but not in the region beyond. 
Otherwise, the \emph{long-chord} regime is well described in terms of Airy functions \cite{ZO}, 
describing an oscillatory region in a ring surrounding the origin, through an outer caustic
and on to an asymptotic evanescent regime as $|\Vxi|\to\infty$ . 
This Airy function results from the uniform approximation 
based on the stationary points of the exponent in \eqref{X-WKB}. 
As mentioned in the Introduction, the stationary points are geometrically identified 
by the intersections between the supporting manifold 
and its translation (\emph{see} fig. \ref{pe}). 

There are two basic reasons why the standard uniform approximation technique,
involving a transformation of the integral \eqref{chi-OVS} into a simpler one (cf. \cite{Berry76,Dingle}),
cannot be employed near the origin. One is that there are an infinite number of stationary points 
of the exponent \eqref{X-WKB} at $\Vxi=0$, whence the origin is a non-generic caustic 
(in the sense of the Thom's classification theorem \cite{Berry76,Thom}). 
On top of this, the large parameter condition in the exponent, 
essential for asymptotic expansions of integral \cite{Dingle}, 
is not fulfilled for the case of small values of $|\Vxi|$, 
where the behavior of the integral is given by \eqref{chi-OVS}, because $|\Vxi|/\hbar\sim1$. 

In order to describe the transition regime (between short and long chords), 
we propose the following (semiclassical) expression for the chord function:
\be
\chi_{sc}(\Vxi)=\chi_s(\Vxi)-SP[\chi_s](\Vxi)+SP[\chi_w](\Vxi).
\label{laformula}
\ee
Here $\chi_s$ is the integral \eqref{chi-OVS}, 
$\chi_w$ is the full semiclassical integral for the chord function, \eqref{super} and \eqref{X-WKB},
and $SP[\cdot]$ denotes the approximation of $[\cdot]$ by stationary phase. 
Notice that $\chi_s(\Vxi)$ was derived as a short chord approximation to $\chi_w(\Vxi)$,
but, since $\Vxi \wedge \x(\theta)$ in the exponent of \eqref{chi-OVS} 
is a nonlinear function of the integration variable, $\theta$,
this integral can also be evaluated by stationary phase.
Indeed, the pair of integrals in \eqref{laformula} that are evaluated by stationary phase 
must cancel in the neighborhood of the origin. On the other hand,
the middle term, $SP[\chi_s]$, cancels $\chi_s$ for large values of $|\Vxi|$, 
where stationary phase evaluation is valid.

The stationary phase method assumes that the integral is dominated by points where the phase is stationary. In the case of the chord function, such stationary points have a geometrical interpretation. Namely, each stationary point defines pairs of points $\x_{\pm}$ on the curve. The point $\x_-$ is the intersection of the classical curve with its translation by the vector $-\Vxi$, whereas $\x_+={\x_-}+\Vxi$. This fitting of the chord into the curve is called a \emph{chord realization}. The stationary points of the chord function are the $q$-coordinates of the centers $\x_j$ of such realizations. In fig. \ref{pe}$b$ we show that each chord has two {\it realizations} in a convex closed curve.
\newline
The amplitude in the above semiclassical approximation may be expressed in terms of the canonical action variable, $I(\x)$. Specifically, the amplitude is given by $| \{I^+, I^- \}|^{-\frac12}$, where $I^{\pm}$ is the action variable at the tips of the realization of $\Vxi$.
Finally, the phase in this approximation is determined by the area $\mathcal{A}_{\x_j}$, between the realization of $\Vxi$ and the curve plus the product $\x_j\wedge\Vxi$. An additional phase, $\sigma_j$, is given by the sign of the expression
\begin{equation}
\frac{\partial^2}{\partial Q^2}[S(\x_j+\Vxi/2)-S(\x_j-\Vxi/2)].
\end{equation}
Therefore the stationary phase evaluation for $\chi_w$ \cite{OVS}, is
\be
SP[\chi_w](\Vxi)=
\frac1{2\pi}\sum_j
\frac{\exp[\frac{i}\hbar\mathcal{A}_{\x_j}(\Vxi)+\x_j\wedge\Vxi+\frac\pi4(\sigma_j(\Vxi)+\gamma_j)]}
{|\{I_j^+,I_j^-\}|^{\frac12}},
\label{SPw}
\ee
where $j$ enumerates the stationary points $\x_j$ and $\gamma_j$ is the difference between the Maslov corrections in \eqref{X-WKB}. 
\par
On the other hand, for $\chi_s$ we have
\be
SP[\chi_s](\Vxi)=\frac{1}{2\pi}\sum_\ell
\frac{[\x''(\theta_\ell(\Vxi))\wedge\Vxi]^{-\frac12}}{\sqrt{2\pi\hbar}}
\exp\left[\frac{i}{\hbar}\x(\theta_\ell(\Vxi))\wedge\Vxi+\frac\pi4\textrm{sign }(\x''(\theta_\ell(\Vxi))\wedge\Vxi)\right],
\label{SPshort}
\ee
where $\ell$ enumerates the stationary points for $\chi_s$, that is, 
the points where the vector $\Vxi$ is tangent to the curve. 
Thus, inserting the explicit stationary phase evaluations \eqref{SPw} and \eqref{SPshort} into \eqref{laformula}
leads to a general approximation, in which the only integral left to be evaluated numerically 
was already present in the short chord approximation \eqref{chi-OVS}.

\section{Nonlinear evolution of a Fock state}

In order to test the general approximation \eqref{laformula}, we analize the chord function 
of a one-parameter set of states, evolving under the action of a simple cubic Hamiltonian,
depending only on momenta:
\be
\label{Hamiltoniano}
H(p)=\alpha_3p^3+\alpha_2p^2+\alpha_1 p+\alpha_0.
\ee
The classical evolution is then determined by
\be
p(t)=p(0)\quad\textrm{ and }\quad q(t)=q(0)+(3\alpha_3p^2+\alpha_2p+\alpha_1)t.
\ee
For a fixed time parameter $t\geq0$, the action function is
$I(p,q,t)=(q-[3\alpha_3p^2+2\alpha_2p+\alpha_1]t)^2+p^2,$
so that the classical curve supporting the evolved state corresponds to the level curve $I(p,q,t)=\II$.

\begin{figure}[htb!]
\centering
\includegraphics[width=15cm]{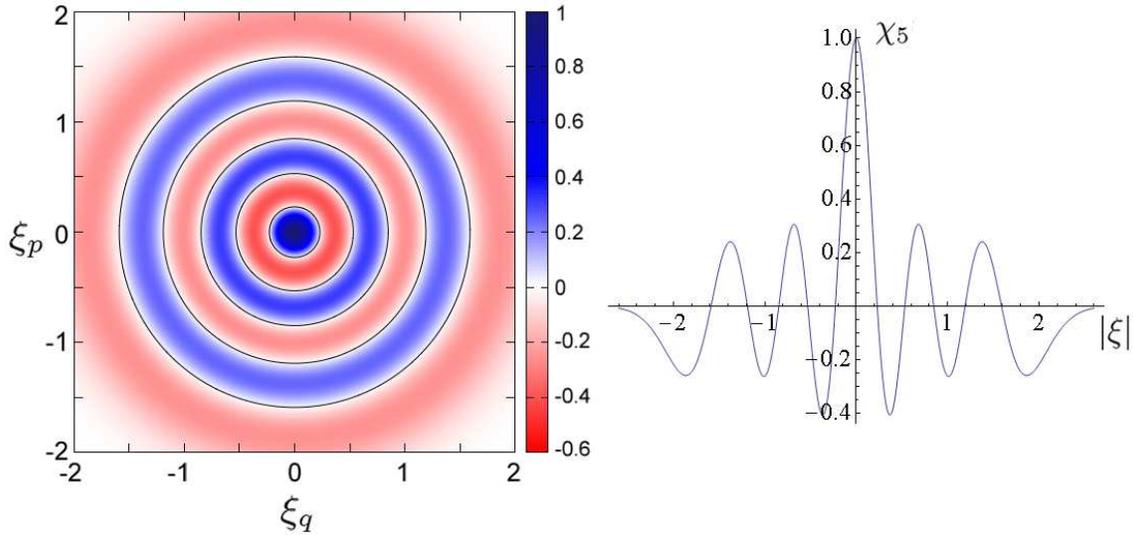}
\caption{(\emph{Left}) Chord function for the $5$th Fock state. It has reflection symmetry and its five nodal lines correspond to circles. (\emph{Right}) Radial cut of $\chi_5(\Vxi)$. Here $\hbar=0.1$\label{fock5}}
\end{figure}

For $t=0$, we choose a Fock state, $|n\rangle$, which is an excited state of the harmonic oscillator of frequency $\omega$. Choosing $\omega=1$, Fock states are supported by a circular manifold, so they are symmetric with respect to reflections. By choosing the center of the supporting circle as the origin, we obtain the (exact) real chord function \cite{OVS},
\begin{equation}
\chi_n(\Vxi)=e^{-|\Vxi|^2/4\hbar}L_n\left(\frac{\Vxi^2}{\hbar}\right),
\end{equation} 
where $L_n(\cdot)$ is the $n$th Laguerre polynomial \cite{Abra}. Since $\chi_n$ is real, it has full nodal lines corresponding to circles, and its radii are $\hbar$ times to the roots of the $n$th Laguerre polynomial, as shown in fig. \ref{fock5}.
The above classical evolution breaks the original central symmetry, because of the cubic term in the Hamiltonian.
For these states, the approximation for the first nodal line, eq. \eqref{ellipse}, provides a circle of radius $\sqrt{\hbar/(n+\frac12)}$. This result differs to the exact with $\sim18\%$ of accuracy; so as previously mentioned, the approximation \eqref{ellipse} just gives a qualitative estimation for the first nodal line.    

 \par
On the other hand, the exact evolving chord function, obtained from \eqref{chitrans}, is
\be
\chi(\Vxi,t)= \langle\psi(t)|\hat{T}_{-\Vxi}|\psi(t)\rangle
= \int dp\,\psi^*_n(p_+)e^{-\frac{i}{\hbar}(t[H(p_+)-H(p_-)]-p\xi_q)}\psi_n(p_-),
\ee
where $p_\pm\equiv p\pm\xi_p/2$ and
\be
\psi_n(p)=\frac{(-i)^n}{\sqrt{2^nn!}}\left(\frac{1}{\pi\hbar}\right)^{\frac14}
H_n\left(\frac{p}{\sqrt{\hbar}}\right)e^{-\frac{p^2}{2\hbar}}
\ee
is the Fock state $n$ in the $p$-representation and $H_n(\cdot)$ is the $n$th Hermite polynomial \cite{Abra}.
\par
\begin{figure}[t!]
\centering
\includegraphics[width=16cm]{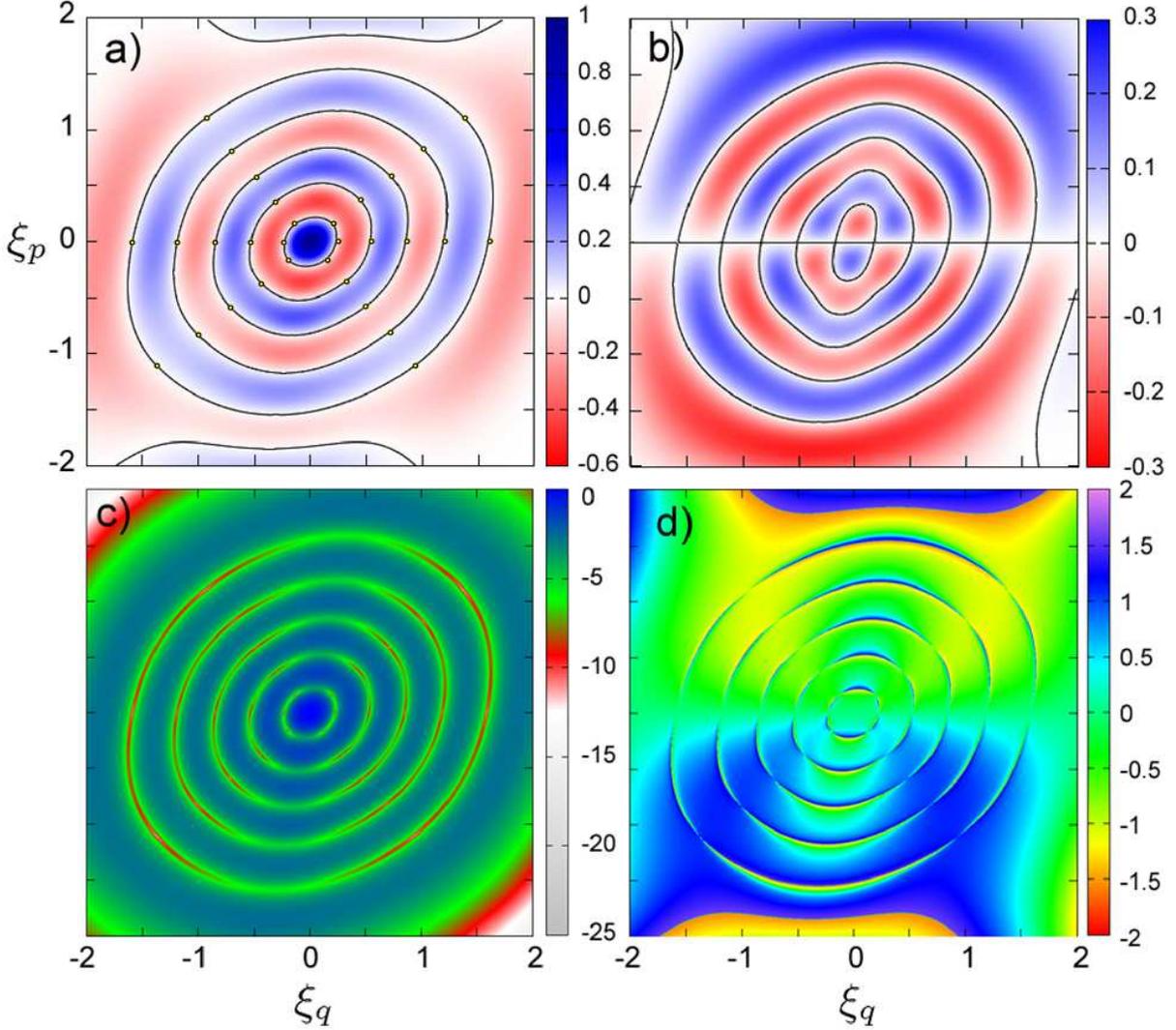}
\caption{\label{fig-chi} The chord function for the evolved Fock state $e^{-i\hat Ht/\hbar}|n=5\rangle$. 
At this resolution, the difference between the exact and the semiclassical approximation is indiscernible. 
The black lines in the real $a)$ and imaginary $b)$ parts are the levels curves for $\chi(\Vxi)= 0$. 
$c)$ is the logarithmic intensity and $d)$ is the phase. The points in $a)$ show the location of the `blind spots'. Here $\hbar=0.1$, $t=0.1$, $\alpha_2=\alpha_1=1$.}
\end{figure}
For any $t>0$, the reflection symmetry is broken, 
thus the chord function has no more a global parity, which implies that the imaginary part is not null. 
As shown in the figures \ref{fig-chi}, the expression \eqref{laformula} describes appropriately 
the behavior of the chord function for short and long chords. Cleary $SP[\chi_w]$ is not valid 
when $\Vxi$ tends to a diameter, but the resulting singularity is avoided 
by using the uniform approximation \cite{ZO}. 
We can recognize the real (fig \ref{fig-chi}a) and imaginary (fig \ref{fig-chi}b) parts 
by their even and odd parity, respectively. The intersections of nodal lines 
for both the real and the imaginary parts correspond to the blind spots, 
i.e. the zeroes of the intensity (local maxima in fig \ref{fig-chi}c) 
and singuralities in the phase (fig \ref{fig-chi}d).
The pair of {\it closest blind spots}, which were approximately specified by the first and second order
moments in section 2, result from the intersection of the smallest closed curve in fig \ref{fig-chi}a
with the straight line through the origin in fig \ref{fig-chi}b. It is curious that the other intersections also seems to imply radial straight lines in this example.

\begin{figure}[htb!]
\centering
\includegraphics[width=8.5cm]{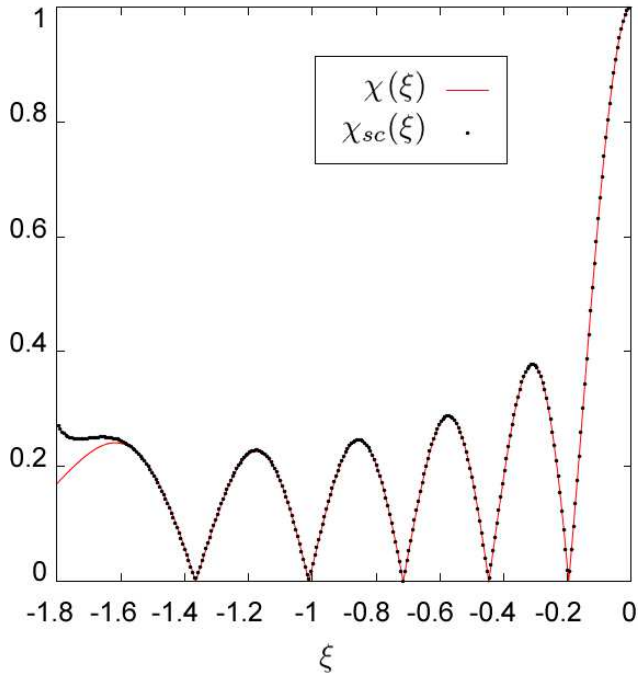}
\caption{\label{fig-zeros}Comparision between the exact and the semiclassical intensities of the chord function for a evolved Fock state $e^{-i\hat Ht/\hbar}|5\rangle$, along the line $\xi_p=0.8172\xi_q$. We can observe a sequence of blind spots.  Here $\hbar=0.1$, $t=0.1$.}
\end{figure}

The detailed comparison of the intensities for our semiclassical approximation of the chord function 
\eqref{laformula} with the exact result is shown in fig \ref{fig-zeros}. The particular radial straight line chosen exhibits a sequence of blind spots. We observe a caustic (the diameter singularity) at the left edge of the figure. It arises in the stationary phase approximation, but this spurious divergence is corrected by the uniform approximation in \cite{ZO}.


\section{Discussion}

The chord function portrays a pure state by exhibiting its overlap with all its possible translations.
There are two cases where such a state has a clear classical correspondence, the superposition of well
separated coherent or squeezed states treated in \cite{ZO} and the Bohr-quantized states analyzed here.
Contrary to naive considerations, in neither case is the decay in the square modulus of the overlap smooth 
and classical like, within a Planck area of the chord origin. We have here shown that
blind spots, denoting zero overlap, arise deep within this classically small neighbourhood.
This feature depends basically on the Schr\"odinger covariance matrix: The greater its determinant,
the closer to the origin will the blind spots lie.  

Knowledge of a finite number of moments determines the chord function of a Bohr-quantized state near the origin, 
but it is insufficient to follow through the complex oscillations,
punctuated by a complex pattern of blind spots, up to the outer limit of an evanescent region. 
We have presented a new semiclassical approximation for the chord function 
of Bohr-quantized states and verified that it is accurate, right up to the outer caustic, 
which was previously treated in \cite{ZO}. In the absence of reflection symmetry, such as present in a Fock state, the chord function is fully complex, which leads to richer structure 
than that displayed by the corresponding Wigner function.
\par
If the system is in contact with an uncontrolled environment, the state will not remain pure, so it must be described by the density operator $\hat\rho$ and the definition \eqref{chitrans} becomes $\chi(\xi)=\tr\hat\rho\hat T_{-\xi}$. Though this chord function still supplies a complete description of the state, the overlap with its translation is now given by
\begin{equation}
C(\xi)\equiv\tr [\ro (\hat T_\xi\ro\hat T^\dag_\xi)]=\frac{1}{2\pi\hbar}\int |\chi(\eta)|^2e^{-i\xi\wedge\eta/\hbar}d\eta.
\end{equation}
For small displacements, $\xi\to0$,  the correlation has a maximal value, since $C(0)=\tr \ro^2=1$. As the displacement increases, we attain an oscillatory regime, where the stationary phase approximation takes account. This behavior changes after the diameter caustic is reached, followed by an evanescent region. For the pure states, this correlation is invariant under Fourier transform, since it reduces to $C(\xi)\sim|\langle\psi|\hat T_\xi|\psi\rangle|^2=|\chi(\xi)|^2$, according to \eqref{invariance}.
\par
As in the case of superpositions of Gaussians wavepackets \cite{Blind}, we expect that blind spots do survive in the chord function for a markovian quantum evolution appropriate to an open system. However they should disappear from $C(\xi)$ much more quickly than the negative regions of the Wigner function. Blind spots correspond to sharp indentations on a background of maximal correlations which makes them measurable. As shown in \cite{ZO}, they take their place as very sensitive indicators of the full quantum coherence alongside the zeroes of the Husimi function \cite{TosAlm99}. The latter often has its zeros in shallow evanescent regions \cite{TosAlm99}, where they are tricky to distinguish.

\acknowledgements
We thank partial financial support by CNPq, FAPERJ, INCT-IQ Informa\c{c}\~ao Qu\^antica (brazilian agencies) and CAPES/COFECUB.

\end{document}